\documentclass[aps,twocolumn,aps,showpacs]{revtex4}

\usepackage[dvips]{graphicx}

\begin{document}

\title{Precision study of 6p $^{2}P_{j}$ $\rightarrow$ 8s $^{2}S_{1/2}$ relative transition matrix elements in atomic Cs}
\author{A. Sieradzan}
\affiliation{Physics Department, Central Michigan University Mt.
Pleasant, MI 48859}
\author{M.D. Havey}
\affiliation{Department of \ Physics, Old Dominion University, Norfolk, VA 23529}
\author{M. S. Safronova}
\affiliation{Electron and Optical Physics Division, National
Institute of Standards and Technology, U.S. Department of
Commerce, Gaithersburg, MD 20899-8410}

\date{\today }

\begin{abstract}
A combined experimental and theoretical study of transition matrix
elements of the 6p $^{2}P_{j}$ $\rightarrow$ 8s $^{2}S_{1/2}$
transition in atomic Cs is reported.  Measurements of the
polarization-dependent two-photon excitation spectrum associated
with the transition were made in an approximately 200 cm$^{-1}$
range on the low frequency side of the 6s $^{2}S_{1/2}$
$\rightarrow$ 6p $^{2}P_{3/2}$ resonance. The measurements depend
parametrically on the relative transition matrix elements, but
also are sensitive to far-off-resonance 6s $^{2}S_{1/2}$
$\rightarrow$ np $^{2}P_{j}$ $\rightarrow$ 8s $^{2}S_{1/2}$
transitions.  In the past, this dependence has yielded a
generalized sum rule, the value of which is dependent on sums of
relative two-photon transition matrix elements.  In the present
case, best available determinations from other experiments are
combined with theoretical matrix elements to extract the ratio of
transition matrix elements for the 6p $^{2}P_{j}$ $\rightarrow$ 8s
$^{2}S_{1/2}$ (j = 1/2,3/2) transition.  The resulting
experimental value of 1.423(2) is in excellent agreement with the
theoretical value, calculated using a relativistic all-order
method, of 1.425(2).
\end{abstract}

 \maketitle


\section{Introduction}
Although some of the earliest experiments in atomic physics were
measurements of atomic lifetimes and oscillator strengths
\cite{molisch, wiese}, precise determination of atomic transition
matrix elements remains a demanding enterprise
\cite{mcalexander,mcalexander2,oates,volz}. Some perspective on
this may be gained by noting that, in spite of the development of
a wide array of sophisticated experimental techniques,
measurements depending directly on atomic transition matrix
elements seldom have achieved a precision better than $\sim$ 0.5
\% this being for the deeply studied alkali atoms
\cite{mcalexander,mcalexander2,oates,volz,volz2,tanner,rafac,young,rafac2,rafac3,vasilyev,ekstrom,jones,wang}.
Measurements made for non-alkali atoms, often motivated by the
need for data in some other area of atomic physics research
\cite{molisch,radzig,cohen,metcalf,mihalas}, have typically cited
even lower precision. We have similarly been concerned with
precise measurements in alkali atoms, and we have a continuing
experimental program of precision measurement of relative and
absolute transition matrix elements in the one-electron atoms
\cite{beger,meyer,havey,bayram,markhotok}. The main experimental
approaches have been based either on polarization-dependent
Rayleigh and Raman scattering in atomic Cs, or on
polarization-dependent two-photon spectroscopy applied to Na and
Rb.

In spite of the fact that many experimental
\cite{tanner,rafac,young,rafac2,rafac3,vasilyev,bennett} and
theoretical \cite{th,li,cs,us,adndt,dzuba} approaches have been
applied to determination of atomic properties of atomic
$^{133}$Cs, including lifetimes or oscillator strengths and
polarizabilities, only limited higher precision experimental data
is available for many valence transition matrix elements in this
atom. These studies have been motivated in part by a serious need
for empirical data to extract more fundamental information from
precise measurement of parity nonconservation (PNC) in this atom
\cite{wood}, especially in light of the discrepancy reported in
\cite{bennett2} between values of the weak charge Qw extracted
from high precision atomic physics experiment \cite{wood} and the
accepted Standard Model value. In order to clarify the situation,
recent experimental efforts have concentrated on measurement of
transition matrix elements associated with the 6s - 6p and 6s - 7p
multiplet transitions \cite{rafac3,vasilyev}.  However, the 6p -
7s and 7p - 7s transitions, which are less experimentally
accessible, make similarly important contributions. The
availability of high-precision experimental data for any
transitions between low-lying states of Cs is important for
providing additional information regarding the accuracy of the
theoretical calculations in Cs which is crucial for the accurate
analysis of Cs PNC experiment.

In this paper we present results of our measurements of the
spectral and polarization dependence of the 6s $^{2}S_{1/2}$
$\rightarrow$ 6p $^{2}P_{j}$ $\rightarrow$ 8s $^{2}S_{1/2}$ (j =
1/2,3/2) two-color, two-photon transition in $^{133}$Cs, where n =
6 is the dominant term. In an earlier report \cite{meyer} we
described how such measurements could be interpreted in terms of a
type of sum rule related to the scalar and vector transition
probabilities.  The sum rule is evaluated by fitting experimental
polarization-dependent spectra to a generalized form containing,
as fitting parameters, the relative two-photon transition matrix
elements for the contributing transitions, allowing extraction of
the ratio of the matrix elements in the dominant term.  The
contributions from far-off-resonance transitions are small but
significant for heavy atoms such as Cs and are evaluated
theoretically. In the present report we describe that approach as
applied to the Cs 6s $^{2}S_{1/2}$ $\rightarrow$ 8s $^{2}S_{1/2}$
two-photon transition.  By combining the present measurements with
precisely determined 6s $^{2}S_{1/2}$ $\rightarrow$ 6p $^{2}P_{j}$
resonance transition matrix elements, the 6p $^{2}P_{j}$
$\rightarrow$ 8s $^{2}S_{1/2}$ relative excited state matrix
elements are determined. We point out that the same combined
scheme can effectively be used to determine matrix elements for
the 6s $^{2}S_{1/2}$ $\rightarrow$ 7p $^{2}P_{j}$ $\rightarrow$ 7,
8s $^{2}S_{1/2}$ transitions which, because of their importance
for analysis of the precision parity nonconservation measurements
in Cs, remain interesting cases to investigate \cite{vasilyev}.

In the following sections, we briefly review our experimental
approach, with particular attention to aspects new in the present
study. This is followed by a description of our experimental
results and data analysis. We then describe the application of
results of relativistic many-body calculations of the
off-resonance terms, which allows the extraction of the desired
two-photon transition matrix element ratios. Finally, the relative
6p $^{2}P_{j}$ $\rightarrow$ 8s $^{2}S_{1/2}$ relative transition
matrix elements are determined and compared with theoretical
calculations.

\section{Experimental Approach}

\begin{figure}
\includegraphics[width=3.0 in]{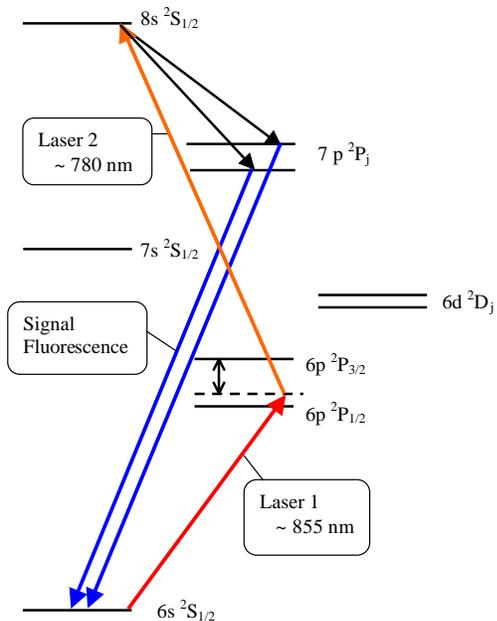}\\
\caption{\label{Figure 1} Partial energy level diagram for atomic
Cs, showing the excitation scheme used in the experiment.}
\end{figure}

The general experimental approach is described in previous reports
\cite{beger,meyer,bayram,markhotok}, and so will only briefly be
reviewed here. The basic experimental scheme \cite{beger} is
illustrated in Figure 1, which contains a partial energy level
diagram for the lowest few levels of atomic Cs. A block diagram of
the experimental apparatus is shown in Figure 2. In the
experiment, atoms in the 6s $^{2}S_{1/2}$ ground level are excited
by two-photon absorption to the 8s $^{2}S_{1/2}$ final level. In
principle, all the intermediate np $^{2}P_{j}$ levels, including
continuum terms, contribute to the total excitation probability
\cite{louden}.  In practice, and depending on the precision of the
measurements, significant contributions are limited to the first
few nearest-to-resonance terms.  The first step of the excitation
scheme is accomplished with an Ar$^{+}$ laser pumped Ti:Sapphire
laser (laser 1) tuned in a several-hundred cm$^{-1}$ energy range
to the red of the 6s $^{2}S_{1/2}$ $\rightarrow$ 6p $^{2}P_{j}$
transition, which has a hyperfine-weighted transition frequency of
$\omega_{3/2}$ = 11732.31 cm$^{-1}$ \cite{moore,udem}. With
reference to Figure 1, the detuning from one-photon resonance is
defined as $\Delta$ = $\omega$$_{1}$ - $\omega$$_{3/2}$, where the
laser 1 frequency is $\omega$$_{1}$. Although detunings of
magnitude greater than 200 cm$^{-1}$ were investigated, useful
data was obtained only for $\mid\Delta\mid$ $<$ 200 cm$^{-1}$. A
Michelson-inteferometer type wavemeter, which has a precision of
10$^{-3}$ cm$^{-1}$, is used to determine $\omega$$_{1}$. The
Ti:Sapphire laser is passively stabilized with a thin-thick etalon
combination and has a short term line width on the order of a few
MHz. Long-term drifts, on a time scale longer than a typical data
run of a few minutes, are dominated mainly by thermal and
mechanical noise, and have a negligible influence on the
experimental results.  The average power is approximately 200 mW.
The laser 1 output is strongly linearly polarized, which is
further purified by passing the beam through a Glan-Thompson prism
polarizer.  The beam is then passed through an electronically
controlled liquid crystal retardation wave plate (LCR), which
switches the linear polarization direction to one of two
orthogonal linear polarization directions.  The resulting variable
polarization beam is then directed to the Cs sample cell.  The
second excitation step is driven by the linearly polarized output
from an external cavity diode laser (ECDL, laser 2), which
generates an average power of 8 mW in a short-term ($\sim$ 1 s)
bandwidth $\sim$ 1 MHz. The laser has a frequency $\omega$$_{2}$
which nominally satisfies the two-photon resonance condition
$\omega$$_{1}$ + $\omega$$_{2}$ = $\omega$$_{0}$, where
$\omega$$_{0}$ is the frequency separation of hyperfine components
of the 6s $^{2}$S$_{1/2}$ and 8s $^{2}$S$_{1/2}$ levels.  The
degeneracy-weighted hyperfine averaged value of $\omega$$_{0}$ =
24317.17 cm$^{-1}$ \cite{moore}.  The laser 2 frequency is
determined from the detuning according to $\omega$$_{2}$ =
$\omega$$_{0}$ - $\omega$$_{3/2}$ - $\Delta$. When $\Delta$ = 0,
the excited state resonance frequency is $\omega$$_{2}$ = 12584.82
cm$^{-1}$ \cite{moore}. The commercial ECDL is piezoelectrically
scannable over a range of approximately 15 GHz around the
two-photon resonance. The diode laser output beam is made to be
nearly collinear with that from the Ti:Sapphire laser and the two
beams are weakly focused with 0.5 m focal length lenses and
overlapped in the central region of a sample cell assembly.

\begin{figure}
\includegraphics[width=3.0 in]{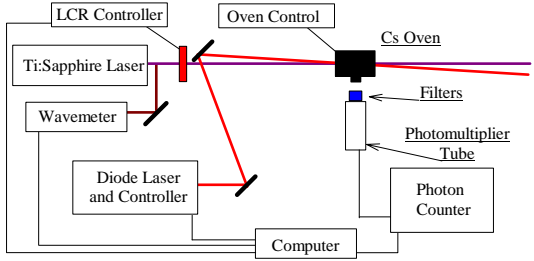}\\
\caption{\label{Figure 2} Schematic diagram of the experimental
apparatus, showing the layout of the main elements of the
experiment.  The Ti:Sapphire Laser is Laser 1, and the Diode Laser
is Laser 2.  The Liquid Crystal Retarder controller is labeled as
the LCR Controller.}
\end{figure}

The Cs vapor cell is an evacuated and sealed Pyrex cell having a
length of about 7.5 cm and a diameter of 2.5 cm. Research grade
windows are fused directly onto the cell body, leaving about a 2.0
cm diameter undistorted central region for transmission of the
laser beams. The cell is prepared on an oil-free vacuum system,
and is evacuated to a base pressure of about 10$^{-8}$ mbar. A
small amount of Cs metal is driven into the cell prior to removing
it from the vacuum system.  The cell temperature is varied by
placing it in a resistively heated oven, which allows heating to a
typical temperature of 375 K, corresponding to a Cs vapor density
of about 2.6 x 10$^{11}$ cm$^{-3}$.  The cell temperature is
stabilized to about $\pm$ 0.1 K using a thermocouple attached to
the coolest part of the cell; the thermocouple output is fed back
to the oven power supply.

Two-photon resonance signals are monitored by measurement of the
7p $^{2}P_{j}$ $\rightarrow$ 6s $^{2}S_{1/2}$ cascade fluorescence
at 455 nm and 459 nm.  The fluorescence is collected at right
angles to the laser beams by a short focal length field lens
($\sim$ 5 cm), which approximately collimates the fluorescence
light on the cathode of a red-light sensitive photomultipler tube
(PMT).  The tube is protected from background light and the
intense laser beams by a combination of colored glass and narrow
band interference filters. Infrared transmitting colored glass
filters mounted on the entrances to the oven housing further
reduce background light signals. The PMT output is amplified and
the photon counting rate measured with a commercial 100 MHz photon
counter.  Typical counting rates on two-photon resonance are
$\sim$ 10$^{3}$ s$^{-1}$.  We point out that the operating
temperature of 375 K turns out to be optimal for this experiment.
At lower temperatures the fluorescence signals are weak, while at
elevated temperatures, and correspondingly larger Cs atom density,
the vapor becomes optically thick to the 7p $^{2}P_{j}$
$\rightarrow$ 6s $^{2}S_{1/2}$ signal radiation.  Then branching
to the 7s $^{2}$S$_{1/2}$ and 6d $^{2}$D$_{j}$ levels (see Figure
1) is enhanced, and signals in the observed decay channel are
correspondingly reduced.

The various instruments are globally controlled or monitored by a
computerized data acquisition and instrument control program.  To
illustrate the experimental protocol, consider a typical
experimental run, where laser 1 is set to a nominal frequency,
which is measured by the wavemeter, which passes this value to the
main control program. Sequentially, the photon-counting rate is
measured alternately for collinear and perpendicular linear
polarization directions of the two lasers. The retardance of the
LCR is controlled with a specially written instrument driver that
communicates with the LCR hardware via the computer parallel port.
The frequency of the ECDL (laser 2) is then shifted by
piezoelectrically scanning the ECDL cavity length; this is
achieved by direct communication between the ECDL controller and
the main experiment computer.  The wavemeter reading is recorded
again and the data cycle is repeated.  This experimental protocol
is very effective for determining the shape of the excitation
line, and for assessing the blending of different hyperfine
components of the excitation line shape.  An alternate data-taking
protocol was also employed. In this, once a proper setting of
laser 2 was determined which minimized the influence of hyperfine
blending (see discussion section) on the measured polarization,
the laser 2 frequency was not scanned. Data was then accumulated
by switching the laser polarization until sufficient statistics
were obtained.

For each data run, the experimental signal is determined for two
different states of relative linear polarization of the excitation
lasers.  Although the absolute intensities of these components
depend on many experimental factors, including the laser
intensities, Cs density, and the sensitivity of the detection
electronics, the intensity ratio is sensitive only to the relative
polarization state of the lasers and the two-photon matrix
elements.  The main experimental observable for detailed analysis
is then the linear polarization degree defined as

\begin{equation}
P_{L} = \frac{S_{\parallel} - S_{\perp}}{S_{\parallel} +
S_{\perp}}, \label{PL}
\end{equation}

where S$_{\parallel}$ and S$_{\perp}$ are the measured signal
intensities when the laser beams are linearly polarized
collinearly or perpendicularly, respectively.

The nuclear spin of $^{133}$Cs is I = 7/2, and so both the 6s
$^{2}S_{1/2}$ and 8s $^{2}S_{1/2}$ electronic states have
hyperfine components of total angular momentum F = 3, 4.  The
hyperfine splitting in the ground 6s $^{2}S_{1/2}$ level is the
International Frequency Standard\cite{bipm}, which is
approximately 9.192 GHz, while the hyperfine splitting in the 8s
$^{2}S_{1/2}$ level is about one-tenth this value and is on the
order of 0.9 GHz. Each of these splittings is larger than the
one-photon Doppler width of several hundred MHz associated with
the separate 6s $^{2}S_{1/2}$ $\rightarrow$ 6p $^{2}P_{j}$ and 6p
$^{2}P_{j}$ $\rightarrow$ 8s $^{2}S_{1/2}$ transitions, and so
partial resolution of the hyperfine splitting is expected even
when the two excitation laser beams are copropagating through the
sample cell.  This is illustrated in Figure 3, which shows the

\begin{figure}
\includegraphics[width=3.0 in]{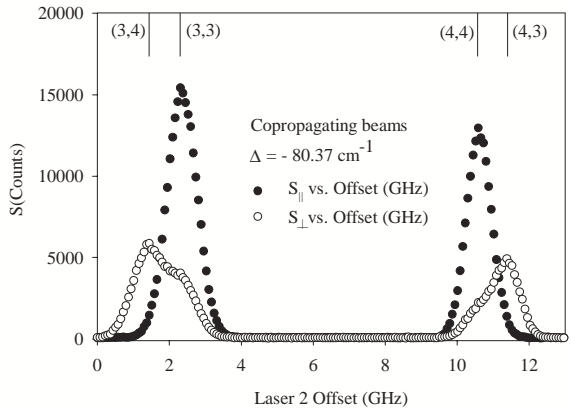}\\
\caption{\label{Figure 3} Experimental signal for frequency and
polarization-dependent two-photon excitation at a detuning of
-80.37 cm$^{-1}$ from resonance.  All four hyperfine transitions
are shown, are notated by (F, F'), and are labeled by the vertical
lines above each component.  The laser 2 offset is measured from a
convenient starting point, denoted by the origin of the graph.
Co-propagating beams.}
\end{figure}

spectral and relative polarization dependence of the two-photon
excitation rate at $\Delta$ = -80.37 cm$^{-1}$.  The main
contribution to the spectral width is due to Doppler broadening,
which is approximately double that for a one-photon transition. On
the other hand, for counterpropagating beams, nearly complete
cancellation of the Doppler width is expected, with a residual
two-photon Doppler width $\sim$ 10 MHz.  A typical such scan is
shown in Figure 4, which corresponds to the F = 4 $\rightarrow$ F'
= 4 transition with $\Delta$ = -22.0 cm$^{-1}$.  There it is seen
that the spectral width of the line shape is approximately 40 MHz,
which corresponds well to the combined residual Doppler width and
the spectral width of the Ti:Sapphire laser.  It is also seen that
the excitation spectrum depends significantly on the relative
polarization state of the two laser beams, giving in this case a
large linear polarization degree of approximately 0.78.

\begin{figure}
\includegraphics[width=3.0 in]{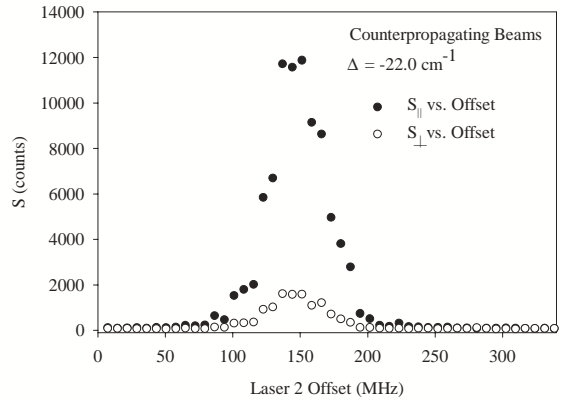}\\
\caption{\label{Figure 4} Experimental signal for frequency and
polarization-dependent two-photon excitation at a detuning of
-22.00 cm$^{-1}$ from resonance. The F = 4 $\rightarrow$ F' = 4
transition is shown. The laser 2 offset is measured from a
convenient starting point, denoted by the origin of the graph.
Counterpropagating laser beams.}
\end{figure}

In spite of the evidently good signal to noise ratio for the data
in Figure 4, it proved difficult to achieve the desired
reproducibility in the polarization measurements with the
passively stabilized Ti:Sapphire laser.  The main reason for this
was that measurement of the two different polarization states were
made sequentially, and fluctuations in the laser frequency on the
switching time scale introduced unwanted noise in the extracted
polarization values.  The reason for this is that when the lasers
are counterpropagating through the interaction region of the cell,
the Doppler width of the two-photon transition is greatly reduced,
and is on the order of the natural width of the final 8s level. In
this case, in order to obtain reliable polarization measurements,
the combined frequency drift of the two lasers would need to be
smaller than a fraction of one MHz/s.  Instead, we recorded the
excitation spectrum for copropagating laser beams, for which the
Doppler width is large (several hundred MHz), and for which the
characteristic laser frequency drift rates of a few MHz/s are
entirely acceptable.  Note that variation of the polarization in
the few MHz range is totally negligible.  The small penalty to be
paid for this is that the polarization values need to be
determined in the wings of the partially blended lines.  However,
these determinations were made sufficiently far into the wings
that the other hyperfine transition made nearly negligible
contribution to the polarization value.  In addition, the signals
were significantly less noisy than with counterpropagating beams,
making possible consistent and repeatable measurements of the
polarization at each detuning.  A typical higher resolution data
run is presented in Figure 5, which corresponds to the F = 4
$\rightarrow$ F' = 3, 4 transitions at $\Delta$ = -167.9
cm$^{-1}$.

\begin{figure}
\includegraphics[width=3.0 in]{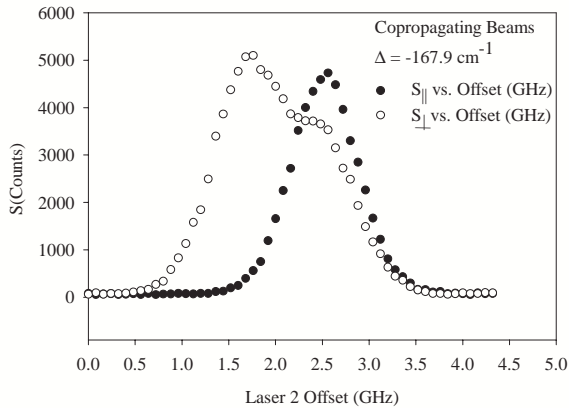}\\
\caption{\label{Figure 5} Experimental signal for frequency and
polarization-dependent two-photon excitation at a detuning of
-167.9 cm$^{-1}$ from resonance. The F = 3 $\rightarrow$ F' = 4
and F = 3 $\rightarrow$ F' = 3 transitions are shown. The laser 2
offset is measured from a convenient starting point, denoted by
the origin of the graph. Copropagating laser beams.}
\end{figure}

Finally, we have modeled the residual effect of blending of the
hyperfine lines on the measured polarization values.  In the
model, we approximate each hyperfine transition line shape to be
Gaussian and having the measured width determined by the Doppler
broadening of the two-photon transition. Using the known relative
hyperfine transition probabilities gives a polarization-dependent
correction that ranges from 0.0005 to 0.0013 (polarization values
reported here can range from -1 to +1).  Even though these
corrections are well within the statistical error associated with
each measured polarization, the corrections are systematic and so
are made directly to the measured counting rates, prior to
extracting P$_{L}$ values from the data through Eq. (1).

\subsection{Analysis and Experimental Results}
The variation of the linear polarization degree with detuning
$\Delta$ of laser 1, and for each of the four electric-dipole
allowed hyperfine transitions, may be readily calculated in terms
of transition matrix elements (the general expression may be found
in \cite{louden}).  The dominant intermediate levels for the
two-photon transitions are the 6p $^{2}P_{j}$ levels, with j =
1/2, 3/2, but other np $^{2}P_{j}$ transitions, including the
p-continuum, also contribute to the total transition probability,
and play an important role in the experiments reported here.  For
the transitions when $\Delta$F = $\pm$1, absorption of two photons
with collinear polarization directions is forbidden, and so the
linear polarization degree is -1, independent of detuning.
However, for the other transition pair, where $\Delta$F = 0, both
fine structure multiplet components contribute, leading to strong
spectral variations in the linear polarization degree. Theoretical
expressions for the intensities are given by

\begin{eqnarray}
\lefteqn{ I_{\parallel} = 4I_{o33}\left[
\frac{R}{\omega_{1}-\omega_{3/2}}+\frac{1}{\omega_{1}-\omega_{1/2}}+
 \right. }\nonumber \\
 && +  \left.
\frac{R}{\omega_{2}-\omega_{3/2}}+\frac{1}{\omega_{2}-\omega_{1/2}}+P\right]^{2}
, \label{I33par}
\end{eqnarray}

\begin{eqnarray}
\lefteqn{ I_{\parallel} = I_{o33}\left[
\frac{R/2}{\omega_{1}-\omega_{3/2}}-\frac{1}{\omega_{1}-\omega_{1/2}}-
 \right. }\nonumber \\
 && -  \left.
\frac{R/2}{\omega_{2}-\omega_{3/2}}+\frac{1}{\omega_{2}-\omega_{1/2}}+Q\right]^{2}
, \label{I33per}
\end{eqnarray}

for the F = 3 $\rightarrow$ F' = 3 transition and

\begin{eqnarray}
\lefteqn{ I_{\parallel} = \frac{36}{15}I_{o44}\left[
\frac{R}{\omega_{1}-\omega_{3/2}}+\frac{1}{\omega_{1}-\omega_{1/2}}+
 \right. }\nonumber \\
 && +  \left.
\frac{R}{\omega_{2}-\omega_{3/2}}+\frac{1}{\omega_{2}-\omega_{1/2}}+P\right]^{2}
, \label{I44par}
\end{eqnarray}

\begin{eqnarray}
\lefteqn{ I_{\parallel} = I_{o44}\left[
\frac{R/2}{\omega_{1}-\omega_{3/2}}-\frac{1}{\omega_{1}-\omega_{1/2}}-
 \right. }\nonumber \\
 && -  \left.
\frac{R/2}{\omega_{2}-\omega_{3/2}}+\frac{1}{\omega_{2}-\omega_{1/2}}+Q\right]^{2}
, \label{I44per}
\end{eqnarray}

for the F = 4 $\rightarrow$ F' = 4 transition.  In these
expressions, the overall intensity for each hyperfine transition
is proportional to I$_{oFF'}$, while R, P and Q are parameters
that are proportional to ratios of reduced transition matrix
elements \cite{rose}. In the present case,

\begin{equation}
R = \frac{<8s\parallel d\parallel6p_{3/2}><6p_{3/2}\parallel
d\parallel6s>}{<8s\parallel d\parallel6p_{1/2}><6p_{1/2}\parallel
d\parallel6s>}, \label{Rdefinition}
\end{equation}

where d is the electric-dipole operator.  The quantities P and Q
are given by

\begin{equation}
P = \sum_{n>6,jk}\frac{M_{njk}}{\omega_{k}-\omega_{np_{j}}}
\label{Pdefinition}
\end{equation}

\begin{equation}
Q=
\sum_{n>6,jk}\frac{(-1)^{j+1/2}}{j+1/2}\frac{M_{njk}}{\omega_{k}-\omega_{np_{j}}},
\label{Qdefinition}
\end{equation}

where the matrix element ratio is given by

\begin{equation}
M_{mjk}=\frac{<8s\parallel d\parallel np_{j}><np_{j}\parallel
d\parallel 6s>}{<8s\parallel d\parallel
6p_{1/2}><6p_{1/2}\parallel d\parallel 6s>}.
\label{Mdefinition}
\end{equation}

The total angular momentum j = 1/2,3/2, while k = 1,2 labels the
frequency of the absorbed photons.

\begin{figure}
\includegraphics[width=3.0 in]{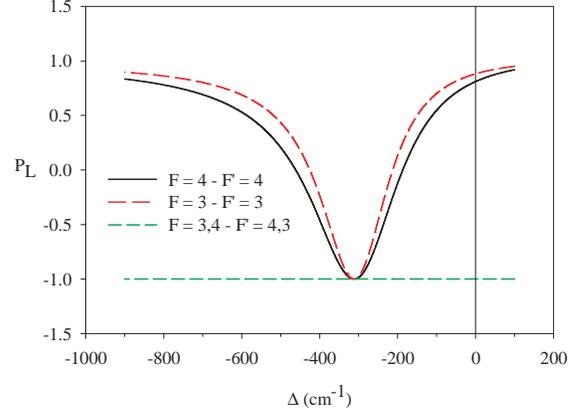}\\
\caption{\label{Figure 6} Model dependence of the variation of the
linear polarization degree with detuning from resonance for all
four permitted hyperfine transitions.  The matrix element ratio is
taken to be R = 2, and contributions from far-off-resonance
transitions are neglected.}
\end{figure}

An illustration of the variation of the polarization with detuning
$\Delta$ is shown in Figure 6 for each of the four hyperfine
transitions. In these plots we have taken R = 2, corresponding to
no relativistic modification of the reduced transition dipole
matrix elements, and P = Q = 0, which applies to the case when
only the n = 6 intermediate levels are considered. Among the
critical features of the plots are the resonance values of the
linear polarization degree for each hyperfine transition. In the
present case, these values are P$_{L}$ = 0.882 for the F = 4
$\rightarrow$ F' = 4 transition, and P$_{L}$ = 0.811 for the F = 3
$\rightarrow$ F' = 3 transition.  In addition, there is a detuning
from resonance where the polarization is P$_{L}$ = - 1.0,
corresponding to the cases when the intensity S$_{\parallel}$ = 0
for each hyperfine transition. The location of this point
generally depends on the four main terms in Eq. (2) - (8), and on
the values for P and Q.

\begin{figure}
\includegraphics[width=3.0 in]{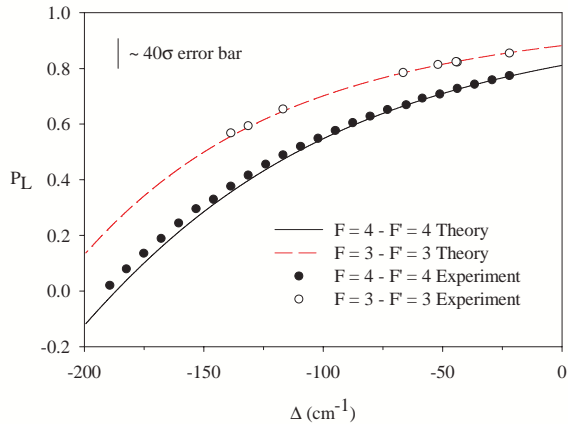}\\
\caption{\label{Figure 7} Measured linear polarization degree of
the two-photon excitation rate as a function of detuning for the F
= 3 $\rightarrow$ F' = 3 and F = 4 $\rightarrow$ F' = 4
transitions. The solid curves represent the model curves for a
matrix element ratio R = 2, and neglecting far-off-resonance
transitions (P=Q=0).}
\end{figure}

Polarization measurements have been made in an approximately 200
cm$^{-1}$ range of detunings to the low frequency side of the
atomic 6s $^{2}$S$_{1/2}$ $\rightarrow$ 6p $^{2}$P$_{j}$ resonance
transition. These measurements, which have a typical one
standard-deviation uncertainty of 0.002, are presented in Figure
7.  The solid curves in this figure are expanded versions of those
in Figure 6.  Note that the error bars on the experimental data
points are negligible on the scale of the figure. It can be seen
from the curves that there is significant discrepancy between the
measurements and the theoretical curves (with P = 0, Q = 0, and R
= 2) for the F = 4 $\rightarrow$ F' = 4 transition, for which we
have taken the most extensive data. At the largest detuning, this
amounts to about 20 standard deviations. The departure has two
main contributions, these being the fact that nonresonant
transitions with n $>$ 6 make a significant contribution through
nonzero values of P and Q, and because the value of R generally
departs from the nonrelativistic value of R = 2.  In previous
work, the three parameters P, Q and R were considered as
constants, which is quite well justified by their weak dependence
on detuning. Fits to the detuning-dependent polarization data then
yielded linear relationships among the parameters; estimates of P
and Q further allowed extraction of a quite precise value for the
constant matrix element ratio R.   Although this is found to be
insufficient for the precision of the data presented here, we
report here our values, making the same assumptions as in earlier
work.  That approach yields R=2.1068(36) + 443.2(1.2)P -
362(362)Q, where the uncertainty (in parenthesis) in the numerical
values represents one standard deviation. Note that although the
coefficient of Q has an uncertainty on the order of its value, its
value is correlated with that of R and P, and so it cannot be
neglected. The final uncertainty in R derived from calculated
values of P and Q shows that the correction due to Q is on the
order of the uncertainty in R. This equation represents the
equivalent sum rule to those presented in earlier reports on our
measurements in Rb.

However, P and Q depend weakly on detuning, and so to obtain the
highest precision, it is desirable to calculate P and Q as a
function of detuning directly from the most accurate experimental
and theoretical matrix elements. We use a combination of the most
reliable experimental and theoretical values for dominant
contributions with n $\leq$ 9.  For n = 6, measured resonance line
(n = 6) matrix elements of Rafac, et al. \cite{rafac3} are
employed, while for the second resonance doublet (n = 7), we use
the recently reported precision measurements of Vasilyev et al.
\cite{vasilyev}. Other values having n $\leq$ 9 are obtained from
the relativistic all-order calculations of Safronova et al.
\cite{us}; these are listed in Table 1.  For multiplets with n $>$
9, which make only a small overall contribution to R, we use
Dirac-Hartree-Fock matrix elements \cite{li}; these values are
also summarized in Table 1, along with the energies associated
with each intermediate np level.  For reference, it is found that
P and Q, calculated with these values, are fit very well by second
order polynomials in detuning $\Delta$ = $\omega$$_{1}$ -
$\omega$$_{10}$, and are given by P = -2.437 x 10$^{-4}$ + 3.257 x
10$^{-9}$$\Delta$ - 3.8847 x 10$^{-12}$$\Delta$$^{2}$ and by Q =
3.260 x 10$^{-6}$ - 7.683 x 10$^{-9}$$\Delta$ + 1.645 x
10$^{-13}$$\Delta$$^{2}$. The detuning $\Delta$ is in vacuum
cm$^{-1}$ units.  We estimate the total uncertainty in P and Q to
be on the order of 1\%.

To obtain the matrix element ratio defined in Eq. (6), a nonlinear
least squares fit of the polarization data of Fig. 7 is made to
the theoretical expressions, using the calculated values of P and
Q for each detuning.  The result of the fit is R = 2.0024(24),
where the dominant error in R comes from the statistical
uncertainty in the measured polarization.  The quality of the fit
is illustrated in Fig. 8.  In the figure, measurements are
compared to the deviation of the polarization from the values
obtained when R = 2, P = Q = 0; these are represented by the
horizontal line passing through zero deviation.  The experimental
measurements, represented by the data points, are seen to be in
excellent agreement with the solid curve, this being calculated
from the fitting parameters.

\begin{figure}
\includegraphics[width=3.0 in]{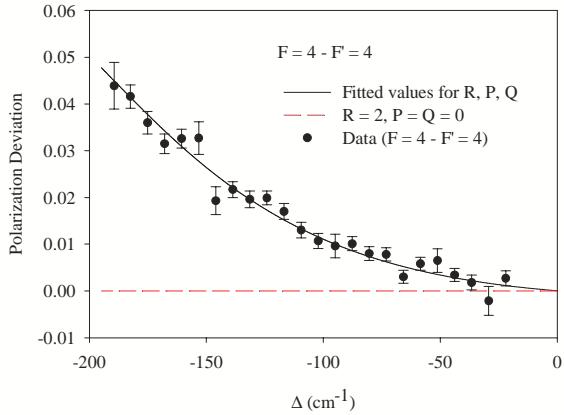}\\
\caption{\label{Figure 8} Deviation of the measured linear
polarization degree of the two-photon excitation rate as a
function of detuning for the F = 4 $\rightarrow$ F' = 4
transition.  The solid curve through the data points represents
the best fit to the experimental data, while the horizontal line
is the model curve result.  Note that the vertical scale is
magnified 20 times in comparison to Figures 6-7.}
\end{figure}

The ratio of excited-state dipole matrix elements R$_{8s-6p}$ =
$<$8s$\parallel$d$\parallel$6p3/2$>$/$<$8s$\parallel$d$\parallel$6p1/2$>$
may be obtained from Eq. (6) by combining the measured value of R
with the high precision 6p resonance line matrix element ratio of
1.4074(3) measured by Rafac et al. \cite{rafac2}. This gives the
excited state matrix element ratio of R$_{8s-6p}$ = 1.423(2),
which has the uncertainty of about 0.15\%, with roughly equal
contributions from statistical uncertainty in the measurements
reported here and from the resonance line matrix element ratio of
\cite{rafac2}.  Finally, we point out that the analysis may be
reversed, if it is assumed that the bare transition matrix element
ratio is calculated precisely.  Then, by recognizing that the
off-resonance terms represented by P and Q contribute an average
of about -0.1091(37) to the value of R, the measurements determine
the much smaller combination of dipole matrix elements in Eqs. (7)
- (8) to about 3 \%. As these terms are dominated by transitions
through the 6s - 7p and 6s - 8p multiplets, the measurements serve
as a consistency check on existing measurements \cite{vasilyev}at
that level.

\section{Relativistic all-order calculation of transition matrix elements}

In order to compare the experimental result $R_{8s-6p}=1.423(2)$
with the high-precision  theoretical value we carry out
relativistic all-order calculation of electric-dipole matrix
elements in Cs. In particular, we calculate $8s-np$, $n=6$, 7, 8,
9 electric-dipole reduced matrix elements using a relativistic
all-order method including single and double (SD) excitations
\cite{li}. The resulting values of the $8s-7p$, $8s-8p$, and
$8s-9p$ electric-dipole matrix elements were used to determine $P$
and $Q$.

 In the SD all-order method, the wave function of the valence electron $v$ is
represented as an expansion
      \begin{eqnarray}
\lefteqn{|\Psi_v \rangle = \left[ 1 + \sum_{ma} \, \rho_{ma}
a^\dagger_m a_a + \frac{1}{2} \sum_{mnab} \rho_{mnab} a^\dagger_m
a^\dagger_n a_b a_a + \right.}\nonumber \\
& \left. \sum_{m\neq v} \rho_{mv} a^\dagger_m a_v + \sum_{mna}
\rho_{mnva} a^\dagger_m a^\dagger_n a_a a_v \right]|\Phi_v\rangle
\label{eq1}
\end{eqnarray}

In this equation $|\Phi_v\rangle$  is the lowest-order atomic
state wave function, which is taken to be the frozen-core
Dirac-Hartree-Fock (DHF) wave function of a state $v$ and
$a^\dagger_i$ and $a_i$ are creation and annihilation operators,
respectively. The indices $a$, $b$ designate core electrons and
indices $m$, $n$ designate any states above the core.
    The equations for excitations coefficients
       $\rho_{ma}$, $\rho_{mv}$,
$\rho_{mnab}$, and $\rho_{mnva}$ are obtained by substituting the
wave function $\Psi_v$ into the  many-body Schr\"{o}dinger
equation
\begin{equation}
H | \Psi_v\rangle=E| \Psi_v\rangle, \label{eq2}
\end{equation}
where $H$ is the relativistic {\em no-pair} Hamiltonian \cite{cs}.
 The equations are solved iteratively
until the corresponding correlation energy for the state $v$ is
sufficiently converged. The resulting excitation coefficients are
then used to calculate matrix elements. The one-body matrix
element of the operator $Z$ given by
      \begin{equation}
      \label{z1}
Z_{wv}=\frac{\langle \Psi_w |Z| \Psi_v \rangle}{\sqrt{\langle
\Psi_v | \Psi_v \rangle \langle \Psi_w | \Psi_w \rangle}}.
\label{eqr}
\end{equation} is expressed in terms of excitation coefficients as
\begin{equation}
 Z_{wv}=\frac {z_{wv}+Z^{(a)}+\cdots +Z^{(t)}} {\sqrt{(1+N_v)(1+N_w)}},
  \label{eqr1}
\end{equation}
where $z_{wv}$ is the lowest-order DHF matrix element, the terms
$Z^{k}$, $k = a...t$ are linear or quadratic functions of the
excitation coefficients, and normalization terms $N_v$ are
quadratic functions of the excitation coefficients. As a result,
certain sets of many-body perturbation theory (MBPT) terms are
summed to all orders. This method is shown to yield high-accuracy
results for the primary transition electric-dipole matrix elements
in alkali-metal atoms \cite{cs,us,th}. The results for the reduced
electric-dipole matrix elements for $8s-6p$, $8s-7p$, $8s-8p$, and
$8s-9p$ transitions are listed in Table~\ref{tab1}.

\begin{table}
 \caption{Energy and matrix elements (ME) used in
calculating the contributions of far-off-resonance transitions to
the measured linear polarization spectrum. ME for n $\geq$ 10 are
obtained from Dirac-Hartree-Fock calculations.}
    \label{tab1}
\begin{tabular}{cccc}
  \hline
\hline
Relay Level & Energy (cm$^{-1}$) & 6s-npj (a.u.) & 8s-npj (a.u.) \\
\hline

6p1/2 & 11178.2 & 4.489$^{a}$ & -1.027$^{d}$ \\
6p3/2 & 11732.4 & 6.324$^{a}$ & -1.462$^{d}$ \\
7p1/2 & 21732.4 & 0.276$^{b}$ & -9.251$^{d}$ \\
7p3/2 & 21765.7 & 0.586$^{b}$ & -14.00$^{d}$ \\
8p1/2 & 25709.1 & 0.081$^{c}$ & 17.71$^{d}$ \\
8p3/2 & 25791.8 & 0.218$^{c}$ & 24.46$^{d}$ \\
9p1/2 & 27637.3 & 0.043$^{b}$ & 1.743$^{d}$ \\
9p3/2 & 27682.0 & 0.127$^{b}$ & 2.969$^{d}$ \\
10p1/2 & 28727.1 & 0.047 & 0.634 \\
10p3/2 & 28753.9 & 0.114 & 1.158 \\
11p1/2 & 29403.7 & 0.034 & 0.348 \\
11p3/2 & 29421.1 & 0.085 & 0.667 \\
12p1/2 & 29852.9 & 0.026 & 0.228 \\
12p3/2 & 29864.7 & 0.067 & 0.451 \\
13p1/2 & 30166.0 & 0.021 & 0.165 \\
13p3/2 & 30174.5 & 0.055 & 0.334 \\
14p1/2 & 30393.2 & 0.017 & 0.127 \\
14p3/2 & 30399.5 & 0.046 & 0.262 \\
15p1/2 & 30563.3 & 0.015 & 0.102 \\
15p3/2 & 30568.0 & 0.039 & 0.213 \\
\hline \hline
\end{tabular}
\begin{flushleft}
$^{a}$Rafac, \emph{et al} \cite{rafac3} \\

$^{b}$Vasilyev, \emph{et al} \cite{vasilyev} \\

$^{c}$Safronova, \emph{et al} \cite{us} \\

$^{d}$This work (all order) \\
\end{flushleft}

 \end{table}

Next, we investigate the effect of the correlation to the $8s-6p$
matrix elements to evaluate the uncertainty of their ratio. We
find that the total correlation correction to the $8s-6p$ matrix
elements is small, 3\%. However, it actually results from severe
cancellations of large contributions illustrated in
Table~\ref{tab2} where we give the breakdown of the all-order SD
calculation for both $8s-6p$ transitions. The lowest-order (DHF)
value is listed in the first row of Table~\ref{tab2}. Three larger
terms, $Z^{(a)}$, $Z^{(c)}$, and $Z^{(d)}$
  \begin{eqnarray}
    Z^{(a)} &=&  \sum_{ma}z_{am} \widetilde{\rho}_{wmva} +
     \sum_{ma}z_{ma} \widetilde{\rho}^{*}_{vmwa} \nonumber \\
    Z^{(c)} &=&  \sum_{m}z_{wm} \rho_{mv} +\sum_{m}z_{mv} \rho^*_{mw}\nonumber \\
    Z^{(d)} &=&  \sum_{mn}z_{mn} \rho^*_{mw} \rho_{nv},
\label{matel}
\end{eqnarray}
where \begin{equation}
   \widetilde{\rho}_{wmva}=\rho_{wmva} - \rho_{wmav}
\end{equation}
are listed separately and all other terms are summed together as
$Z_{\text{other}}$. The total normalization correction
$Z_{\text{norm}}$ defined as
     \begin{equation}
            Z_{\text{norm}}=Z_{wv}-[z_{wv}+Z^{(a)}+\cdots +Z^{(t)}]
\end{equation}
 is given separately. We find that two dominant terms, $Z^{(c)}$ and $Z^{(d)}$, nearly cancel each
other and their sum almost exactly cancels out the term $Z^{(a)}$.

\begin{table}
\begin{ruledtabular}
 \caption{Contributions to SD all-order electric-dipole  $8s-6p$
matrix elements in Cs. Matrix elements are given in a.u.}
 \begin{tabular}{lrr}
  \multicolumn{1}{c}{Contribution}&
  \multicolumn{1}{r}{$8s-6p_{1/2}$}&
  \multicolumn{1}{r}{$8s-6p_{3/2}$}\\
  \hline
    DHF                 &  1.0584 &   1.5145\\
    $Z^{(a)}$           &  0.0173 &   0.0177\\
    $Z^{(c)}$           & -0.1122 &  -0.1606\\
    $Z^{(d)}$           &  0.0938 &   0.1317\\
    $Z_{\mathrm{other}}$& -0.0084 &  -0.0108\\
    $Z_{\mathrm{norm}}$ & -0.0228 &  -0.0308\\
    Total               &  1.0260 &   1.4618\\
  \end{tabular}
  \label{tab2}
\end{ruledtabular}
 \end{table}

We note that while terms $Z^{(a)}$ and $Z^{(c)}$ contain
third-order terms as well as higher-order terms  (see \cite{th}
for details) the $Z^{(d)}$ term, being quadratic in valence
single-excitation coefficients $\rho_{mv}$ contains only terms
starting from fifth order. Out of the remaining terms, the largest
contribution comes from the normalization correction. For these
transitions, the dominant contribution to $N_{v(w)}$ in the
denominator of Eq.~(\ref{eqr1})  comes from the term
    \begin{equation}
 \sum_{m} \rho^*_{mv} \rho_{mv}.
\end{equation}
This term contributes  97\%  to the $N_{8s}$ value  and 64\% to
the $N_{6p}$ value. Again, this term is quadratic in single
excitation coefficients $\rho_{mv}$ and can contain only MBPT
terms starting from fifth order. Thus, $8s-6p$ matrix elements
present an interesting case with large, but canceling,
contributions from high orders in many-body perturbation theory.
This cancellation occurs for both $8s-6p_{1/2}$ and $8s-6p_{3/2}$
matrix elements.

    To evaluate the uncertainty in the theoretical ratio
     $R_{8s-6p}$ we also calculate  $8s-6p_{3/2}$
and $8s-6p_{1/2}$ electric-dipole matrix elements in different
approximations and estimate some omitted contributions. The
results are summarized in Table~\ref{tab3}. The lowest-order DHF
values are listed in column labeled DHF.  The results of the
third-order many-body perturbation-theory calculation, which
includes higher-order random-phase-approximation terms as
described in \cite{adndt}, are listed in column ``III''. The
third-order values, which include an estimate of the omitted
fourth- and higher-order Brueckner-orbital (see classification and
formulas in \cite{adndt}) corrections obtained by the scaling
procedure described in \cite{adndt}, are listed in column
``III$_{sc}$''. Single-double all-order data from Table~\ref{tab1}
are listed in column ``SD''. The results obtained by including
partial contribution of the triple excitations are listed in
column labeled ``SDpT''. These data are obtained by adding a
triple-excitation valence term to Eq.~(\ref{eq1}) and making
corresponding corrections to correlation energy and single-valence
excitation coefficient equations $\rho_{mv}$ as described in
Refs.~\cite{cs,us,th}. The $\rho_{mv}$ excitation coefficients
give rise to the largest terms, $Z^{(c)}$ and $Z^{(d)}$, and the
dominant part of  $N_v$.

All-order SD and SDpT calculations include a complete third-order
contribution (see \cite{th} for a detailed comparison) but omit
some classes of higher-order terms starting from fourth order. We
have estimated some omitted correlation corrections resulting from
triple and higher excitations using the scaling described, for
example, in Ref.~\cite{cs}. Briefly, single-particle excitation
coefficients $\rho_{mv}$ are multiplied by the ratio of the
experimental and theoretical correlation energies. The modified
excitation coefficients are then used to re-calculate matrix
elements. Such scaling estimates only certain classes of the
omitted contributions (mainly Brueckner-orbital contributions, see
Ref.~\cite{cs} for classification of the perturbation theory
terms). The corresponding SD and SDpT scaled values are listed in
columns ``SD$_{sc}$'' and ``SDpT$_{sc}$''. \textit{Ab initio} and
scaled values are grouped together.

  \begin{table}
\begin{ruledtabular}
 \caption{Electric-dipole 8s-6p$_{j}$ reduced matrix elements in Cs calculated using different
approximations: Dirac-Hartree-Fock (DHF), third-order many-body
perturbation theory (III), single-double all-order method (SD),
single-double all-order method including partial triple
contributions (SDpT) and the corresponding scaled values.
$R_{8s-6p}$ is the ratio of the $8s-6p_{3/2}$ and $8s-6p_{1/2}$
matrix elements. Absolute values of the matrix elements in a.u.
are given. }
 \begin{tabular}{lrrrrrrrrrr}
 \multicolumn{1}{c}{Level}&
\multicolumn{2}{c}{}& \multicolumn{3}{c}{\textit{Ab initio}}&
\multicolumn{1}{c}{}&
 \multicolumn{3}{c}{Scaled}\\
 \multicolumn{1}{c}{}&
\multicolumn{1}{c}{DHF}&
 \multicolumn{1}{c}{}&
 \multicolumn{1}{c}{III}&
\multicolumn{1}{c}{SD}&
  \multicolumn{1}{c}{SDpT}&
   \multicolumn{1}{c}{}&
   \multicolumn{1}{c}{ III$_{\mathrm{sc}}$}&
 \multicolumn{1}{c}{ SD$_{\mathrm{sc}}$}&
\multicolumn{1}{c}{ SDpT$_{\mathrm{sc}}$}  \\
           \hline
    j=1/2& 1.0584&&  1.0231 &   1.0260&  1.0321&&   1.0315& 1.0223& 1.0327 \\
    j=3/2& 1.5145&&  1.4519 &   1.4618&  1.4709&&   1.4712& 1.4556& 1.4705 \\
    $R_{8s-6p}        $& 1.4309&&  1.4191 &   1.4247&  1.4252&&   1.4262& 1.4238& 1.4240 \\
  \end{tabular}
  \label{tab3}
\end{ruledtabular}
 \end{table}
 
        The third-order matrix elements do not differ significantly from the all-order results
indicating very accurate cancellation between large higher-order
terms. The Brueckner- orbital correction is dominant in a
third-order calculation for both transitions and is relatively
more important for the $8s-6p_{3/2}$ transition, thus giving a
ratio differing from more accurate SD all-order value by 0.6\%. We
note that the ratio of the scaled third-order values, which
includes an estimate of higher-order Brueckner-orbital terms,
agrees with the all-order result. As one can see from
Table~\ref{tab3}, the total correlation correction contribution to
the ratio $R_{8s-6p}$ is very small, 0.4\%, owing to the
cancellation of dominant terms. The difference between all-order
data in different approximations is only 0.1\%.  As the largest
contributions to these matrix elements contain valence
single-excitation coefficients $\rho_{mv}$ the inclusion of the
partial triple contributions and the scaling described above
should give a good estimate of the omitted higher-order terms as
both these methods are aimed at correcting $\rho_{mv}$.  We
recommend the SD \textit{ab initio} value $R_{8s-6p}=1.425(2)$ as
the final theoretical value for the ratio. The uncertainty is
obtained by combining the uncertainty of the dominant terms,
determined to be 0.1\% based on the spread of SD, SDpT and scaled
values, and the total uncertainty in all other, much smaller,
contributions taken to be 0.1\% not to exceed the uncertainty of
the dominant terms.

\section{Discussion of results and conclusions}
Comparison of the theoretical matrix element ratio of R =
1.425(2), and the corresponding experimental result of R =
1.423(2) shows excellent agreement.  This is remarkable, given the
various contributing theoretical and experimental factors that
significantly affect the final result in each case.  It is
interesting to reiterate that many of the two-photon transition
matrix elements in heavy alkali atoms show significant
relativistic modification. For example, the 5s $^{2}S_{1/2}$
$\rightarrow$ 5p $^{2}P_{j}$ $\rightarrow$ 5d $^{2}D_{3/2}$
transition in Rb shows such variations at a level of nearly 7 \%.
At the same time, other ratios, such as that measured in the
present case, show significantly smaller relativistic
modification, and yet in each case such modifications are
generally well-described by the highest-level atomic structure
calculations.  Nevertheless, there is a significant sensitivity in
the experimental measurements to other dipole transitions, and
these transitions depend significantly on relativistic
contributions. Naturally, it would be of interest to measure the
absolute oscillator strengths of the individual multiplet
transitions, in order to make direct comparison of the matrix
element values reported here. Such measurements remain a
significant challenge to experimental technique. Finally, the
excited state transitions in atomic Cs associated with the 7p
doublet are of particular interest, as those second resonance line
matrix elements show nearly a factor of 2 departure from the
expected line strength ratio of 2.

In conclusion, we have used precision two-photon polarization
spectroscopy to make measurements of the transition matrix element
ratio associated with the 6s $^{2}S_{1/2}$ $\rightarrow$ np
$^{2}P_{j}$ $\rightarrow$ 8s $^{2}S_{1/2}$ transition in atomic
Cs. The measurements are combined with other experimental data and
calculations in order to extract the ratio of excited-state matrix
elements.  The experimental value is found to be in excellent
agreement with the ratio of reduced matrix elements calculated
using a relativistic all-order method.

\begin{acknowledgments}
The financial support of the National Science Foundation
(NSF-PHY-0099587) and Central Michigan University is greatly
appreciated.
\end{acknowledgments}

\end{document}